\documentstyle[emulateapj]{article}

\newcommand\mmax{$\langle m_{\rm max}\rangle$}
\newcommand\mmaxx{m_{\rm max}}
\newcommand\mup{$m_{\rm up}$}

\newcommand\pmmaxmup{$p(\mmaxx | m_{\rm up})$}

\newcommand\msol{\rm\,M_\odot}

\def\spose#1{\hbox to 0pt{#1\hss}}
\def\simpropto{\mathrel{\spose{\raise 3pt\hbox{$\propto$}}
     \lower 3.0pt\hbox{$\sim$}}}

\received{15 September 2004}
\accepted{}

\slugcomment{Accepted 23-Dec-2004 to the Astrophysical Journal Letters}

\lefthead{M. S. Oey \& C. J. Clarke}
\righthead{Stellar Upper Mass Limit}

\begin{document}

\title{
Statistical Confirmation of a Stellar Upper Mass Limit
}
\author{M. S. Oey}
\affil{University of Michigan, Department of Astronomy, 830 Dennison
	Building, Ann Arbor, MI\ \ \ 48109-1090, USA}
\author{\medskip and \\ C. J. Clarke}
\affil{Institute of Astronomy, Madingley Road, Cambridge CB3 0HA, UK}

\begin{abstract}
We derive the expectation value for the maximum stellar mass ($\mmaxx$)
in an ensemble of $N$ stars, as a function of the IMF upper-mass cutoff
(\mup) and $N$.  We statistically demonstrate that the upper IMF of
the local massive star census observed thus far in the Milky Way and
Magellanic Clouds clearly exhibits a universal upper mass cutoff
around $120 - 200\ \msol$ for a Salpeter IMF, although the result is
more ambiguous for a steeper IMF.  
\end{abstract}


\section{Introduction}

The upper mass limit to the stellar initial mass function (IMF) is a
critical parameter in understanding stellar populations, star
formation, and massive star feedback in galaxies.  To date, the
largest empirical mass estimate for an individual star is around $200
- 250\ \msol$ for the Pistol Star (Figer et al. 1998) near the
Galactic Center, and around $120 - 200\ \msol$ for the most massive
stars in the Large Magellanic Cloud (e.g., Massey \& Hunter 1998).  In
practice, most  
applications assume an upper mass limit to the IMF of $m_{\rm up}\sim
100$ to $150\ \msol$.  However, there is some confusion
on whether the apparent observed upper limit simply represents a
statistical limit owing to a lack of sampled stars in individual
clusters (Massey 2003; Massey \& Hunter 1998; Elmegreen 1997).  

Before the advent of the {\sl Hubble Space Telescope} ({\sl HST}),
stars with extremely high masses $\gtrsim 1000\ \msol$ were suggested
to exist.  The dense stellar knot R136a in the 30 Doradus star-forming
region of the LMC was 
the best-known candidate for harboring such a star (Cassinelli,
Mathis, \& Savage 1981).  The viable
candidates for these supermassive stars were eventually resolved by
{\sl HST} and ground-based imaging into smaller stars within the
conventionally observed mass range (e.g., Weigelt et al. 1991;
Heydari-Malayeri, Remy, \& Magain 1988).  In recent years,
however, the possibility of supermassive stars is receiving renewed
attention as a possible mode of star formation in the early universe
(e.g., Bond, Arnett, \& Carr 1984; Larson 1998; Bromm, Kudritzki \&
Loeb 2001).  

It is therefore important to clarify expectations for the highest-mass
stars compared to the existing observations.  Elmegreen (2000)
quantitatively demonstrates that, in the absence of an upper-mass
cutoff, stellar masses should be observed up to $40,000\ \msol$ for
the entire Milky Way, based on estimates for the current star
formation rate and molecular gas mass.  Here, we derive the behavior
of the expectation values for the most massive stars and demonstrate
that, for a universal IMF, current observations indeed show the
existence of an upper-mass limit around $m_{\rm up}\sim 120 - 200\ \msol$.

\section{The Expectation Value \mmax}

\begin{figure*}
\epsscale{1.0}
\plotone{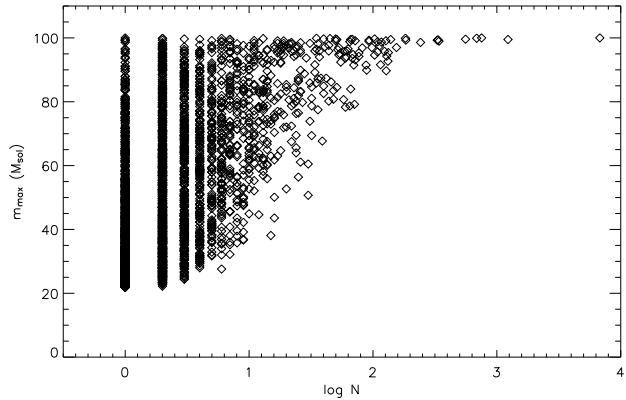}
\vspace*{-2.0in}
\caption{Monte Carlo simulation showing the maximum stellar mass
$m_{\rm max}$ per cluster vs the number of stars $\log N$ per cluster
for 5000 clusters in a distribution of $N$ given by
equation~\ref{eq_nstar}.  A Salpeter IMF is adopted with stellar
masses between 20 -- 100 $\msol$ in this simulation.
\label{f_montecarlo} }
\end{figure*}

Because of the decreasing power law form of the IMF, the
characteristic mass of the largest star formed in clusters of $N$
stars decreases as $N$ decreases.  Figure~\ref{f_montecarlo}
demonstrates this effect with a Monte Carlo simulation.  $N$ is drawn
for individual star clusters from the universal power-law distribution
in $N$ (e.g., Oey \& Clarke 1998; Elmegreen \& Efremov 1997):
\begin{equation}\label{eq_nstar}
n(N)\ dN \propto N^{-2}\ dN \quad ,
\end{equation}
and the stellar masses for each cluster of $N$ stars is drawn from the
Salpeter (1955) IMF, within a mass range of 20 to 100 $\msol$:
\begin{equation}\label{eq_imf}
\phi(m)\ dm \propto m^{-2.35}\ dm \quad .
\end{equation}
Figure~\ref{f_montecarlo} shows the distribution of the most massive
star in each cluster, $m_{\rm max}$ vs $\log N$.  For single stars, we
confirm that the bin of $\log N=0$ is described simply by the IMF
(equation~\ref{eq_imf}).  It is apparent that for large $N$, one can
expect that $m_{\rm max}\simeq m_{\rm up}$, but that for small $N$,
the typical most massive star is much lower in mass.  For $N=1$,
the typical $m_{\rm max}$ is the mean of the IMF, which is 37 $\msol$
for the distribution of $20\leq m\leq 100\ \msol$ used in
Figure~\ref{f_montecarlo}. 

We can analytically derive the expectation value \mmax\ for the most
massive star in an ensemble of $N$ stars as follows.  For $N$ stars,
the probability that all are in the mass range 0 to $M$ is,
\begin{equation}
P(0,M) = \Biggl[\int_{0}^M \phi(m)\ dm\Biggr]^N \quad ,
\end{equation}
where $\phi(m)$ corresponds to the IMF, i.e., a probability distribution
function whose integral is unity.  It follows that the probability
that all the stars are in the mass range 0 to $M+dM$ is, 
\begin{equation}
P(0,M+dM) \simeq \Biggl[\int_{0}^M \phi(m)\ dm\Biggr]^N +
	\frac{d}{dM}\Biggl[\int_{0}^M \phi(m)\ dm\Biggr]^N \ dM
\end{equation}
by Taylor expansion.  Thus we see that the probability that the
most massive star is in the range $M$ to $M+dM$ is,
\begin{equation}\label{eq_prob}
P(M,M+dM) = 
   \frac{d}{dM}\Biggl[\int_{0}^M \phi(m)\ dm\Biggr]^N \ dM \quad ,
\end{equation}
and the expectation value for the most massive star is,
\begin{equation}
\langle m_{\rm max}\rangle = \int_{0}^{m_{\rm up}} M\ 
   \frac{d}{dM}\Biggl[\int_{0}^M \phi(m)\ dm\Biggr]^N \ dM \quad .
\end{equation}
Integrating by parts, this yields,
\begin{equation}\label{eq_mmax}
\langle m_{\rm max}\rangle = m_{\rm up} - 
   \int_{0}^{m_{\rm up}} \Biggl[\int_{0}^M \phi(m)\ dm\Biggr]^N \ dM \quad .
\end{equation}
For large $N$, equation~\ref{eq_mmax} confirms that $\langle m_{\rm
max}\rangle\rightarrow m_{\rm up}$, corresponding to an IMF that is
well-sampled up to the upper mass limit (termed ``saturated'' by Oey
\& Clarke 1998). 

\begin{figure*}
\epsscale{1.0}
\plotone{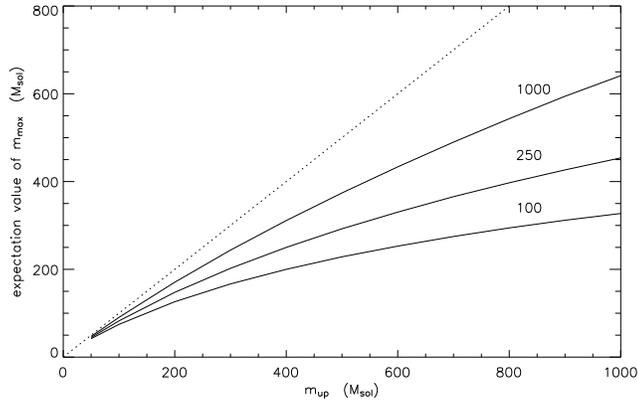}
\vspace*{-2.0in}
\caption{The expectation value \mmax\ vs upper mass limit \mup, for
$N=100,\ 250$, and 1000 stars having masses above $m_{\rm lo} = 10\
\msol$, assuming a Salpeter IMF.  The dotted line shows $m_{\rm
max}=m_{\rm up}$ for comparison. 
\label{f_mmax} }
\end{figure*}

We numerically integrate equation~\ref{eq_mmax} using a lower mass limit
$m_{\rm lo} = 10\ \msol$ instead of 0, and assuming the Salpeter
IMF.  Figure~\ref{f_mmax} shows the
expectation value for the most massive star \mmax\ vs the upper mass
limit \mup\ for $N=100,$ 250, and 1000 stars (solid lines).  The
dotted line shows the identical relation $\langle m_{\rm
max}\rangle=m_{\rm up}$ for comparison.  For lower $N$, \mmax\ is
smaller at any given \mup, as expected. 

\section{Results}

\subsection{R136a}

We start by comparing Figure~\ref{f_mmax} to the R136a region in 30
Doradus, which, at an age of 1--2 Myr (Massey \& Hunter 1998), is
sufficiently young that none of its stars have expired yet as
supernovae.  We consider stars having $m > 10\ \msol$, of which
Hunter et al. (1997) found $N=650$ in this region.  We however note
that this value represents a strong lower limit, since the star
counts are significantly incomplete between 10 and 15 $M_\odot$ (Massey
\& Hunter 1998).  Figure~\ref{f_mmax} demonstrates that the
expectation value of $\mmaxx$ is  
considerably greater than the observed maximum of  $\sim120-200\ \msol$
in R136a, unless \mup\ is low ($\ll 500\ \msol$).  Weidner \&
Kroupa (2004) reached the same conclusion from a similar analysis of 
R136a; Selman et al. (1999) also suggested a cutoff using a less
rigorous analysis.  We can furthermore assess the statistical significance of
this result by calculating \pmmaxmup, the probability 
of obtaining an observed maximum stellar mass $\leq \mmaxx$ for a
given \mup.  Table~\ref{t_sample} lists \pmmaxmup\ calculated from
equation~\ref{eq_prob} for R136a for a range of \mup.  This
demonstrates the negligible likelihood ($ < 10^{-5}$) 
that R136a is drawn from a population which extends to 1000 $\msol$.   
 
The results of Weidner and Kroupa (2004), and those presented here, 
do not support the suggestion by Massey (2003) and Massey \& Hunter
(1998) that the upper IMF in the 
R136a is consistent with $m_{\rm up}= \infty$.  Massey \& Hunter (1998)
found that the penultimate mass bin in the empirical mass function
is fully consistent with the Salpeter slope.  However, they omit from
their mass function, and from their analysis, stars with inferred
masses $> 120\ \msol$, because the lack of stellar models in the grid
preclude reliable mass determinations.  For the two effective temperature
scales they adopted, there are 2 or 9 of these omitted, most-massive
stars.  Although we do not know the exact masses, their 
numbers are sufficient to determine whether an upper mass cutoff to
the IMF power law exists.  For a Salpeter IMF, in the absence of an upper
mass cutoff, there should be a total of 1.7 times more stars at
$m> 120\ \msol$ than are found in the the mass bin 85 -- 120$ \msol$.
Massey \& Hunter (1998) count (8, 11) stars in the latter mass bin,
therefore implying that (14, 19) stars should be found at higher
masses.  This is significantly more than the (2, 9) stars found.
Thus, R136a exhibits a cutoff around $120 - 200\ \msol$,
consistent with the finding by Weidner \& Kroupa (2004).

\subsection{A sample of young OB associations}

Although a truncated IMF in R136a seems conclusively demonstrated,
it is possible that the dense, rich cluster environment of this region
represents a special case.  Can we draw a similar conclusion
from a wider sample of ordinary OB associations?  To examine this further,
%
we consider the upper IMF from the substantial sample of OB
associations that have been uniformly studied by Massey and
collaborators, who estimated stellar masses from spectroscopic
classifications.  Massey, Johnson, \& DeGioia-Eastwood (1995) tabulate
the numbers of stars having $m\geq 10\ \msol$ in the Milky Way and LMC
associations.  To minimize the possibility that the most massive
stars have already expired as supernovae, we count only stars in OB
associations with ages $\leq 3$ Myr.  
Table~\ref{t_sample} shows the observed $N(\geq 10\msol)$ and
$\mmaxx$ for these objects.

We now compute \pmmaxmup\ for all the objects (Table~\ref{t_sample}).  These 
show that, although none of these regions individually provide strong
constraints on the upper mass cutoff, they collectively point to a
conclusion similar to that found for R136a.  The total $N=263$ stars, 
for which inspection of Figure~\ref{f_mmax} again
shows that the observed maximum stellar masses imply that \mup\ should
not exceed a few hundred $\msol$.  Elmegreen (2000) reached a similar
conclusion based on the lack of supermassive stars in the entire
population of the Milky Way.  In considering the total of 263
stars, or Milky Way population, we assume that the IMF is a universal
probability distribution function that is independent of specific
conditions in individual clusters and parent molecular clouds.
Indeed, the IMF is conventionally treated as a universal function
(see, e.g., Elmegreen 2000).  We also emphasize that our total counts
of $N$ are conservative lower limits, since additional young massive
stars can be counted from associations studied by other authors.  We
chose not to include these additional stars in the interest of
maintaining a uniform and well-understood sample.

\begin{figure*}
\epsscale{1.0}
\vspace*{2.0in}
\hspace*{-1.5in}
\plotone{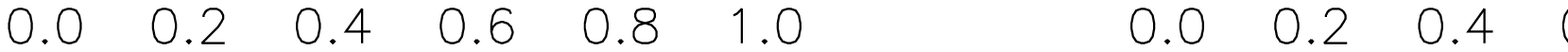}
\vspace*{-4.5in}
\caption{Distribution of \pmmaxmup\ for the sample of OB
associations in Table~\ref{t_sample}, adopting $\mmaxx=120\ \msol$ for
R136a and Tr~14/16.  The aggregate probability that
these distributions originate from a uniform distribution are
$P < 0.002,\ < 0.02,\ < 0.12,$ and $< 0.47$ for, respectively, \mup\ $=
10^4,\ 200,\ 150,$ and 120 $\msol$.
\label{f_centhist120} }
\end{figure*}

\begin{figure*}
\epsscale{1.0}
\vspace*{3.5in}
\hspace*{-1.5in}
\plotone{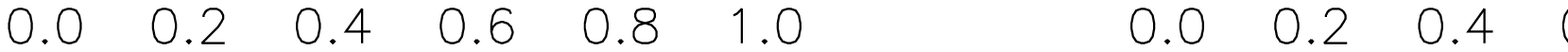}
\vspace*{-4.0in}
\caption{Same as Figure~\ref{f_centhist120}, but adopting an observed
$\mmaxx$ for R136a and Tr~14/16 of $200\ \msol$.  The aggregate
probability that these distributions originate from a uniform
distribution between 0 and 1 are $P < 0.002,\ < 0.002,\ < 0.47,$ and
$< 0.92$ for, respectively, \mup\ $= 10^4,\ 10^3,\ 200,$ and 150 $\msol$.
\label{f_centhist200} }
\end{figure*}

Furthermore, we can now evaluate the total probabilities $P$ that
the values of \pmmaxmup\ represent uniform distributions
between 0 and 1, as expected for any universal \mup.  For example, we
would expect 10\% of the regions to fall into the category where 
\pmmaxmup\ was $\leq 0.1$, 20\% to have a
\pmmaxmup\ of $\leq 0.2$, and so on.  Figure~\ref{f_centhist120}
shows, for each assumed \mup\ of the parent IMF, the distribution of
\pmmaxmup\ for the individual regions.  It is evident that for higher
values of \mup, the values of \pmmaxmup\ are unacceptably clustered
towards small values.  A K-S test confirms this conclusion, yielding
probabilities that these values are uniformly distributed, of
$P < 0.002,\ < 0.02,\ < 0.12,$ and $< 0.47$
for, respectively, \mup\ $= 10^4,\ 200,\ 150,$ and 120 $\msol$.  
We also compute $P$ for an adopted observed $\mmaxx =  200 \msol$,
as might be possible for Tr~14/16 and R136a (Table~\ref{t_sample}).
Figure~\ref{f_centhist200} shows the respective results in this case:
$P < 0.002,\ < 0.002,\ 0.47,$ and $< 0.92$ for \mup\ $= 10^4,\ 10^3,\ 200,$ 
and 150 $\msol$.  We therefore see that \mup\ $=\infty$, and even
$10^3\ \msol$, are effectively ruled out.
{\it Hence the results from this wider total sample of OB
associations points to an upper-mass limit to the IMF around the
observed values of $120 - 200\ \msol$.} 

\begin{deluxetable}{lccllcccc}
\footnotesize
\tablewidth{0pt}
\tablecolumns{7} 
\tablecaption{Sample of OB Associations \label{t_sample}} 
\tablehead{ \colhead{Name} &
\colhead{$N(>10\ \msol)$} & \colhead{$\mmaxx$} & \colhead{$p(10^4)$} &
\colhead{$p(10^3)$} & \colhead{$p(200)$} & \colhead{$p(150)$} & \colhead{$p(120)$}
 }
\startdata
R136a \tablenotemark{a}& 650 & 120 & $10^{-10}$ & $10^{-10}$ & $10^{-5}$ & 0.002 & 1.000 \\
R136a \tablenotemark{b} & 650 & 200 & $10^{-5}$ & $10^{-5}$ & 1.000  &  \nodata  &  \nodata  \\
Berkeley 86 & 10  &  40  &  0.188 & 0.192  &   0.224  &  0.244  &  0.268 \\
NGC 7380    & 11  &  65  &  0.400 & 0.409  &   0.486  &  0.534  &  0.592 \\
IC 1805     & 24  &  100  &  0.335 & 0.350 &  0.510  &  0.626  &  0.784 \\
NGC 1893    & 19  &  65  &  0.206 & 0.213 &   0.288  &  0.338  &  0.404 \\
NGC 2244    & 12  &  70  &  0.407 & 0.416 &   0.502  &  0.556  &  0.623  \\
Tr 14/16 \tablenotemark{a}& 82 & 120 &  0.055 & 0.064 &  0.231  &  0.464  &  1.000 \\
Tr 14/16 \tablenotemark{b}& 82 & 200  &  0.236 & 0.276 & 1.000 & \nodata & \nodata \\
LH 10       & 65  &  90  &  0.032 & 0.037 &   0.102  &  0.176  &  0.324 \\
LH 117/118  & 40  &  100  &  0.161 & 0.174 &  0.326  &  0.458  &  0.666  \\
\enddata
\tablenotetext{a}{Values obtained by adopting 120 $\msol$ for the most
	massive observed stars.}
\tablenotetext{a}{Values obtained by adopting 200 $\msol$ for the most
	massive observed stars.}
\end{deluxetable}

\section{Conclusion}

We have analyzed the upper IMF in a sample of young, nearby OB
associations that best represents stellar census data in this regime.
The clusters are young enough that their highest-mass
members remain present, and the stellar masses are spectroscopically
determined by Massey and collaborators (Massey et al. 1995; Massey \&
Hunter 1998).  Our results provide clear evidence for an upper
truncation in the IMF.   While this result
has been previously noted by Weidner \& Kroupa in the
case of R136a, we show here that it also applies
to a much wider sample of OB associations.  We have furthermore
quantified the statistical significance of such statements.
For example, we find that the probability that the stellar
population of R136a is drawn from a parent distribution
having \mup$= 10^4\ \msol$ is $< 10^{-5}$, and for other
associations the probability is only a few percent.

It should be noted that our results are sensitive to the slope of the
IMF for stars more massive than $10\ \msol$.  In this mass range, it
is often reported that the IMF power-law exponent is close
to the Salpeter value $\sim -2.35$  (e.g., Massey 2003; Schaerer 2003;
Kroupa 2002).  However, should the slope through some systematic
observational bias be significantly steeper, then our 
demonstration of an upper-mass cutoff becomes less vivid.  For
example, we find that adopting an IMF slope of $-2.8$ yields an
aggregate probability that the clusters originate from an IMF having
\mup$=10^4\ \msol$ of $P<0.30$, contrasted to $P<0.002$ for the
Salpeter slope.  Conversely, for a parent IMF slope flatter than the 
Salpeter value, the existence of an upper-mass cutoff is even more
strongly demonstrated.  For R136a, $p(10^4) = 6\times 10^{-4}$ for the
steeper slope ({\it cf.} Table~1), which is still a negligible
probability.  Weidner \& Kroupa (2004) examine the influence of the
slope in more detail.


Thus, given the standard Salpeter slope for massive stars,
it is hard to escape the conclusion that the IMF is truncated
near $m_{\rm up}\sim 120 - 200\ \msol$, based on this analysis.
If these results are real, the only other possibilities are that the
IMF is not universal, or there is an extreme selection effect that
prevents our observations of the most massive stars.
We note that \mup\ need not be an absolute limit, but represents at
least a dramatic drop from the power-law form of the IMF.  
Our conclusion depends on the assumption that the highest
mass stars have not already expired, and it therefore depends
critically on the reliability of evolutionary models for the most
massive stars, and on the reliability with which one can assign ages
to OB associations.  It also assumes that the stars in the OB
associations are coeval.  Should the star
formation process indeed be suppressed at high masses, as suggested by
our results, a major goal for theorists will be to identify
the physics, e.g., plausibly associated with stellar feedback, that
introduces this mass scale into the star formation process.


\acknowledgments
We are pleased to acknowledge discussions with Phil Massey and Don Figer.
We also thank the referee, Rolf Kudritzki, for useful comments.


\end{document}